# The Effect of *In vivo*-like Synaptic Inputs on Stellate Cells

**Dongwook Kim**

Department of Mathematics, Atlanta Metropolitan State College, Atlanta, Georgia, 30310 USA

## Abstract

Previous experimental work has shown high-frequency Poisson-distributed trains of combined excitatory and inhibitory conductance- and current-based synaptic inputs reduce amplitude of subthreshold oscillations of SCs. In this paper, we investigate the mechanism underlying these phenomena in the context of the model. More specially, we studied the effects of both conductance- and current-based synaptic inputs at various maximal conductance values on a SC model. Our numerical simulations show that conductance-based synaptic inputs reduce the amplitude of SC's subthreshold oscillations for low enough value of the maximal synaptic conductance value but amplify these oscillations at a higher range. These results contrast with the experimental results.

## 1. Introduction

Neural activity often oscillates at specific frequencies or frequency bands. Neural oscillations are ranged from very slow oscillations with periods of minutes to very fast oscillations with frequencies reaching 600 Hz [1]. Those oscillations are believed to be involved in processing of visual and spatial information at multiple time scales [2-5]. Frequency bands can be detected by electroencephalography (EEG) recordings which are the recording of electrical activity produced by the firing of neurons within the brain. One of the most studied rhythms is the theta rhythm (4-12 Hz) [4]. Theta frequency oscillations have been observed in various area of brain including the entorhinal cortex and the hippocampus [4]. When a rat is engaged in active motor behavior such as walking or exploratory sniffing, and during REM (dreaming) sleep [6, 7]. In addition, theta frequency oscillations are believed to be involved in spatial learning, memory and navigation [3, 4]. Other frequency bands are labeled the delta (1-4 Hz) (known as slow-wave sleep and characterizing the depth of sleep), beta (13-30 Hz) (normal waking consciousness) and gamma (30-70 Hz) frequency (closely associated with sensory processing and creating the unity of conscious perception) [2].

In this paper, we focus on theta rhythmic activity in the medial entorhinal cortex (MEC). The MEC is the interface between the neocortex and the hippocampus. Studies of cortical connectivity have shown that the superficial layers (II and III) of the MEC receive convergent inputs from various neocortices and these convergent information is delivered to the hippocampus via the perforant path [8, 9]. which provides a connectional route from the entorhinal cortex to various areas on the hippocampal formation, including the dentate gyrus, CA1, CA3, and the subiculum. Conversely, the hippocampus feeds back onto layer V of the MEC and this information flows back from the MEC to the neocortices. Electrophysiological behavior of the MEC is characterized by the presence of robust rhythmic activity in the theta frequency band [10]. In layer II, it is generated primarily by so called stellate cells (SCs) which are the most abundant cell type and the main component of the perforant path. *In vitro* experiments and theoretical studies using biophysical (conductance-based) models [7, 11] have shown that SCs have the intrinsic and dynamic properties that endow them with the ability to display rhythmic activity in the theta frequency band which persist during synaptic transmission block, as originally demonstrated by [12]. More clearly, SCs display rhythmic subthreshold membrane potential oscillations in the theta frequency range and, when the membrane is set positive to threshold, SCs fire action potentials at the peak of the subthreshold oscillations, but not necessary at every subthreshold oscillation cycle [7]. These subthreshold oscillations have been shown to result from the interaction between a persistent sodium and a hyperpolarization-activated (h-) currents [7, 11]. With different levels of injected $I_{app}$ current, SCs show subthreshold oscillations, action potentials and coexistence of subthreshold oscillations and action potentials. Because of these intrinsically rhythmic properties, SCs may play a role in the generation of theta oscillations in the hippocampus [6]. It has been suggested that theta oscillations create the appropriate temporal dynamics between presynaptic activity and post-synaptic excitability that favors synaptic plasticity [4]. And it has also been proposed that the MEC might contribute to its memory functions through synchronizing mechanisms [13] by which the activity patterns of multiple cortical inputs that converge on MEC neurons may be temporally coordinated for the production of a memory representation [4, 14]. For these reasons, the investigation and the understanding of the mechanisms underlying processing information are of great importance [15].

Previous experimental work has shown that medial entorhinal cortex (MEC) layer II stellate cells (SCs) exhibit subthreshold oscillations and resonance in the theta frequency band [16-18]. These intrinsic properties of SCs play an important role in the activity of neural networks in the entorhinal cortex and hippocampus [4]. An interesting question is under what conditions subthreshold oscillations on SCs can be generated at the theta frequency band in the presence of *in vivo*-like synaptic inputs. Fernandez et al. [19] have shown that subthreshold oscillations in SCs are reduced under high-frequency Poisson-distributed trains of combined excitatory and inhibitory conductance-based synaptic inputs while these oscillations does not change under current-based synaptic inputs. Here, we investigated the mechanism underlying these phenomena in the context of the model. More specifically, we studied the effects of both conductance- and current-based synaptic inputs at various maximal synaptic conductance values on a SC model. We found that conductance-based synaptic inputs reduce subthreshold oscillations for low enough value of the maximal synaptic conductance value but amplify these oscillations at a higher range. This is in contrast to the experimental results [19].

## 2. Model

### 2. 1 Biophysical models for stellate cell

We use a single-compartment biophysical (conductance-based) neuron model, introduced by [15], which is based on measurement from layer II stellate cell (SC) of the medial entorhinal cortex (MEC) [7, 20, 21]. This model has the standard action potential producing Hodgkin-Huxley sodium ($I_{Na}$), potassium ($I_K$) and leak current ($I_L$). And two additional currents, a persistent current ($I_P$) and a hyperpolarization-activated, mixed cation current called the h-current ($I_h$) [7], that have been shown to be responsible for the generation of subthreshold oscillations [11]. This model has been used to explain the synchronization properties of strongly coupled excitatory SCs [7]. The current balance equation for single SCs is

$$C\frac{dV}{dt} = I_{app} - I_{Na} - I_K - I_L - I_h - I_p \quad (1)$$

where $V$ is the membrane potential ($mV$), C is the membrane capacitance ($\mu F/cm^2$), $I_{app}$ is the applied bias (DC) current ($\mu A/cm^2$), $I_{Na} = G_{Na}m^3h(V_s - E_{Na})$, $I_K = G_K n^4(V_s - E_K)$, $I_L = G_L(V_s - E_L)$, $I_h = G_h(0.65r_f + 0.36r_s)(V_s - E_h)$, and $I_p = G_p(V_s - E_{Na})$. In Eq. (1), $G_x$ and $E_x$ ($x = $ Na, K, L, p, h) are the maximal conductances ($mS/cm^2$) and reversal potential ($mV$) respectively. The units of time are *msec*. All the gating variables $x$ ($x = $ m, h, n, p, $r_f$, $r_s$) obey a first order differential equation of the following form:

$$\frac{dx}{dt} = \frac{x_\infty(V)}{\alpha_x(V)+\beta_x(V)} \quad (2)$$

where the activation and inactivation curves and voltage-dependent time scales are given respectively by

$$x_\infty = \frac{\alpha_x(V)}{\alpha_x(V)+\beta_x(V)} \, , \, \tau_x(V) = \frac{1}{\alpha_x(V)+\beta_x(V)} \quad (3)$$

Unless stated otherwise, we use the following parameter values $E_{Na} = 55$, $E_K = -90$, $E_L = -65$, $E_h = -20$, $G_{Na} = 52$, $G_K = 11$, $G_L = 0.5$, $G_p = 0.5$, $G_h = 1.5$ and C = 52. The definition of $\alpha_x$ and $\beta_x$ are given in Appendix.

## 2.2 Channel white noise in persistent sodium channel

Many modeling studies have introduced channel noise to obtain robust subthreshold oscillations (STOs) [20, 21, 22]. White et al [20] showed that the number of persistent $Na^+$ channels underlying STOs is relatively small, and argued that the stochastic behavior of these channels may contribute crucially to the cellular level responses. Fransen et al. [21] used a noisy model having $I_p$ and a two component $I_h$. They concluded that, although noise is not required for the SC to display STOs, its presence increases their robustness [21]. We introduce channel white noise in the persistent sodium current ($I_p$)[11, 20]. More specifically, we added a following stochastic term $\xi$ to the dynamic equation for $p$.

$$\frac{dp}{dt} = \frac{p_\infty - p + \xi}{\tau_p} \quad (4)$$

Where $\xi$ represents Gaussian noise with zero mean value and variance appropriate for equilibrium conditions at the current value of membrane potential. For a population of N channels and a time step *dt*, $\xi$ can be implemented by adding the term

$$\xi = \sqrt{\frac{-2dt(\alpha_p(1-p)+\beta_p p)\ln(r_1)}{N}} \cos(2\pi r_2)$$

to the gating equation at each time step [22, 23]. In this equation, $r_1$ and $r_2$ are pseudo-random numbers drawn from a uniform distribution over (0,1].

## 2.3 Biophysical reduced 3D NAS SC Model

In Rotstein et al. [11], they on it studied the mechanism of generation of subthreshold oscillations and the onset of spikes in the SC model (Eq. 1). Using dimensionality reduction methods, they uncovered a three-dimensional model that captures the dynamics of the SC in the subthreshold voltage regime [11] where subthreshold oscillations are generated and the onset of spikes occur. This reduced model provides a good approximation to the full SC model in that regime. This reduced 3D model describes the evolution of the membrane potential V and the two *h*-current gating variables $r_f$ and $r_s$ (fast and slow respectively). Notably, $I_{Na}$ and $I_K$ were found to have a negligible effect on subthreshold dynamics and thus, were omitted from the reduced model [11]. The persistent sodium gating variable p evolves much faster than the remaining gating variables and the adiabatic approximation $p = p_\infty(V)$ was made. The resulting equations are

$$C\frac{dV}{dt} = I_{app} - G_p p_\infty(V)(V - E_{Na}) - I_L - I_h \quad (5)$$

$$\frac{dr_f}{dt} = \frac{r_{f,\infty}(V)-x}{\tau_{r,f}(V)} \quad (6)$$

$$\frac{dr_s}{dt} = \frac{r_{s,\infty}(V)-x}{\tau_{r,s}(V)} \quad (7)$$

Where $I_{p,\infty}(V) = G_p p_\infty(V)(V - E_{Na})$. Eqs. (5) - (7) describe the dynamics of the SC in subthreshold regime, including both the generation of subthreshold oscillations and the onset of spikes [11], but they do not describe the spike dynamics which belong in a different regime (where $I_{Na}$ and $I_K$ are the main active currents). If spiking dynamics are not of interest but only the occurrence of spikes, the dynamics of a SC can be approximately described by Eqs. (5) - (7) supplemented with a zero width artificial spike (occurring on a short time scale and reaching a peak of about 60mV) and an appropriate threshold ($V_{th}$) and reset ($V_{rst}$) values. This reduced model was termed the Nonlinear Artificially Spiking SC (NAS-SC) model [11]. It is a class of model that includes the generalized integrate-and-fire (GIF) and resonate-and-fire models [24].

They choose $V_{th}$ such that it lies close to the end of the subthreshold regime. Once the trajectory reaches the threshold value the voltage is reset to its initial, subthreshold value. Note that in contrast to other models, crossing Vth is not part of the mechanism of spike generation and only indicates spike occurrence [11]. The onset of spikes, however, is accurately described by Eqs. (5) - (7). The reset values $r_f$=0 and $r_s$=0 can be derived from the seven-dimensional SC model [11]. The reset value $V \approx -80\ mV$ is an estimate from our numerical simulations. Thus, the initial conditions in the subthreshold regime is $(V, r_f, r_s) = (-80, 0, 0)$. They reset the trajectory to these values after each spike has occurred. As in the 7D model, we added white Gaussian noise in the persistent sodium channel. To add noise in 3D model, we added the noise terms in $p_\infty$ term in Eq. (4) by $p_\infty + \xi$ where $\xi$ is the Gaussian noise with mean value of zero and variance 1.

## 3. The Effect of *In vivo*-like Synaptic Inputs on Stellate Cells

### 3.1 Spike-Train Statistics

A spike train is a series of discrete action potentials from a neuron taken as a time series. This string of neuronal _ring may be generated spontaneously or as a response to some external stimulus. The stochastic relationship between a stimulus and a response stems from the probabilities corresponding to every sequence of spikes that can be evoked by the stimulus. The probability of two spikes occurring together is not necessary to the product of the probabilities that they occur individually, because the presence of one spike may affect the occurrence of the other. If the spikes are statistically independent, the firing rate is all that is needed to the probabilities for all possible action potential sequences. A stochastic process that generates a series of action potentials is called a point process. The probability of one spike occurring at any given time could depend on the entire history of preceding spikes, so that the intervals between successive spikes are independent, the point process is called a renewal process. If there is no dependence at all on preceding spikes, so that the spikes are statistically independent, we have a Poisson process. The Poisson process provides an extremely useful approximation of stochastic neuronal firing.

### 3.2 Poisson Process

A Poisson process is the simplest possible random process with no memory and is characterized by a single parameter, the rate of mean frequency λ. It is of great relevance to neurobiology, since a number of discrete spikes appear to follow a Poisson distribution closely [21, 22]. Let $\{N(t), t \geq 0\}$ be a sequence of spike-time (random spike times, Poisson process) with mean rate λ if:

1. Given any, the random variables $t_0 < t_1 < t_2 < \cdots < t_{n-1} < t_n$, the random variables $N_{tk} - N_{tk-1}, k = 1, 2, \cdots, n$ are mutually independent.

2. For any $0 \leq t_{k-1} < t_k$ the average number of spikes occurring between $t_{k-1}$ and $t_k$ is $\lambda(t_k - t_{k-1})$

The first condition indicates that the number of spikes occurring in one interval is independent of the number of spikes occurring in any other interval, provided they do not overlap. The second property tells us that the expected number of spikes is proportional to the rate times the duration of the interval.

It follows from these conditions that the actual number of spikes, $N(t_k) - N(t_{k-1})$, is a random variable with the Poisson probability distribution

$$\Pr\{N(t_k) - N(t_{k-1}) = n\} = \frac{(\lambda(t_k - t_{k-1}))^n e^{-\lambda(t_k - t_{k-1})}}{n!} \tag{8}$$

with $n = 0, 1, 2, \cdots$, The parameter $\lambda$ specifies the average number of events per unit time. With $t_{k-1} = t$ and $t_k = t + \Delta t$, we have for the probability that exactly $k$ spikes occur in the interval $\Delta t$.

$$\Pr\{N(t + \Delta t) - N(t) = n\} = \frac{(\lambda \Delta t)^n e^{-\lambda \Delta t}}{n!} \tag{9}$$

If $\lambda \Delta t \ll 1$, that is, if much less than one event is expected to occur in the interval $\Delta t$, the $e^{-\lambda \Delta t}$ term can be expanded into a Taylor series $1 - \lambda \Delta t + \frac{(\lambda \Delta t)^2}{2!} - \frac{(\lambda \Delta t)^3}{3!} + \cdots$ Therefore,

$$\Pr\{N(t + \Delta t) - N(t) = 0\} = e^{-\lambda \Delta t} \approx 1 - \lambda \Delta t \tag{10}$$

$$\Pr\{N(t + \Delta t) - N(t) = 1\} = \lambda \Delta t e^{-\lambda \Delta t} \approx \lambda \Delta t \tag{11}$$

## 3.3 Generating Poisson Spike Trains

There are two commonly used procedures for numerically generating Poisson spike trains The first approach is based on the approximation in equation 4.4 for the probability of a spike occurring during a short time interval. For the homogeneous Poisson process (the rate $\lambda$ is constant), this expression can be rewritten as

$$P\{1 \text{ spike during } \Delta t\} = \lambda \Delta t$$

This equation can be used to generate a Poisson spike train by first subdividing time into a bunch of short intervals, each of duration $\Delta t$. Then generate a sequence of random number $x_i$, uniformly distributed between 0 and 1. For each interval, if $x_i \leq \lambda \Delta t$, generate a spike. Otherwise, no spike is generated.

The second approach for generating a homogeneous Poisson spike train is simply to choose inter spike intervals randomly from the exponential distribution. Each successive spike time is given by the previous spike time plus the randomly drawn inter spike interval. Now each spike is assigned a continuous time value instead of a discrete time bin. However, to do anything with the simulated spike train, it is usually much more convenient to discretely sample the spike train, which makes this approach for generating the spike times equivalent to the first approach described above. Here, we used the second approach for generating Poisson spike train.

## 3.4 Methods

We follow [19]. SCs received inputs consisting of combined excitatory and inhibitory Poisson-driven synaptic conductances or currents. Synaptic inputs consisted of two independent Poisson processes generating unitary synaptic events. Inhibitory and excitatory events were delivered at a frequency of 1000 and 500 *Hz* respectively. The frequency of excitatory and inhibitory inputs was based on the fact that spontaneous background inhibitory frequency is greater than excitatory frequency in slices [27]. In the presence of conductance-based synaptic inputs, subthreshold oscillations are highly attenuated or entirely eliminated. Conversely, with current-based synaptic input stellate cells retain their ability to generated subthreshold oscillation in the theta band [19]. The introduction of conductance-based synaptic inputs abolished the peak at theta frequencies in the power spectrum density (PSD) for subthreshold voltage traces. Under current-based synaptic inputs, which generated a comparable level of membrane voltage fluctuations, the PSD does not change its peak at theta frequencies.

### 3.4.1 Modeling Approach

We follow [19]. Synaptic protocols consisted of two independent Poisson processes generating unitary synaptic events (*Se* and *Si* for excitatory and inhibitory inputs respectively). These were modeled using biexponential functions of the form:

$$S(t) = e^{-\frac{t}{t_{dec}}} - e^{-\frac{t}{t_{rise}}} \tag{12}$$

For GABA$_A$-based inhibition, $t_{rise} = 0.5\ ms$ and $t_{dec} = 5\ ms$. For AMPA-based excitation $t_{rise} = 0.25\ ms$ and $t_{dec} = 2.5\ ms$. [19].

The values of the reversal potentials are $E_e = 0\ mV$ and $E_i = -75\ mV$. For excitation and inhibition, individual synaptic events had the same maximal conductance.

From equation 1, the current-balance equation for both conductance and current-based synaptic inputs is given by

$$C\frac{dV}{dt} = I_{app} - I_{Na} - I_K - I_L - I_h - I_p - I_{syn} \tag{13}$$

where

$$I_{syn} = \sum_{j=1}^{N_e} H(t-t_j) G_e S_e(t_j)(V-E_e) + \sum_{k=1}^{N_i} H(t-t_k) G_i S_i(t_k)(V-E_i) \tag{14}$$

$$I_{syn} = \sum_{j=1}^{N_e} H(t-t_j) G_e S_e(t_j) + \sum_{k=1}^{N_i} H(t-t_k) G_i S_i(t_k) \tag{15}$$

$G_e$ and $G_i$ are maximal synaptic conductance for excitatory and inhibitory inputs respectively and $t_j$ and $t_k$ are independent Poisson distributed times.

Figure 1 illustrates the presynaptic activity of $S_e$ and $S_i$ for excitatory and inhibitory activity respectively. The mean value of $S_i$ is greater than the mean of $S_e$. Note that the mean value of $S_e$ is 0.2704 and the mean value of $S_i$ is 0.4880.

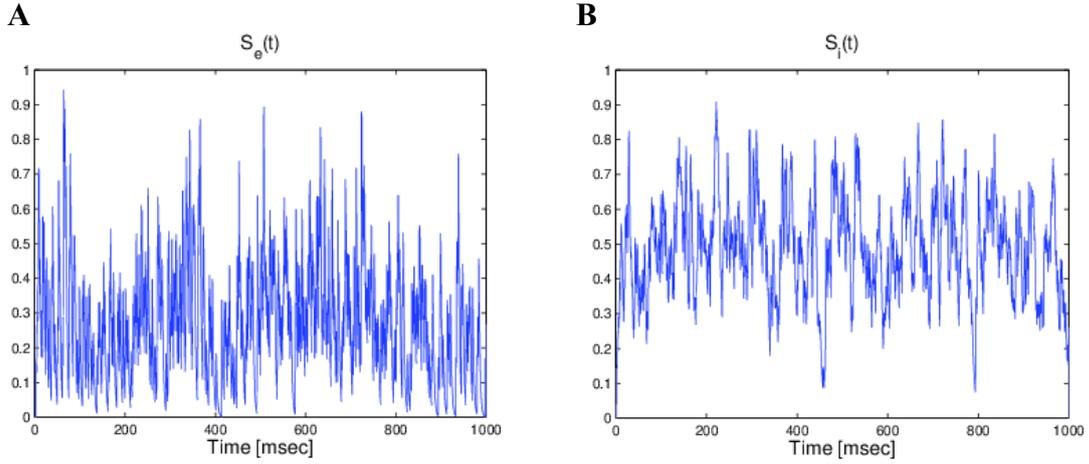

Figure 1: Activity of excitatory (left panel) and inhibitory (right panel) driven Poisson process. The mean value of $S_i$ is almost double to the mean value of $S_e$ : the mean value of $S_i$ and $S_e$ is 0.4880 and 0.2704 respectively.

We begin with a characterization of the power spectrum density (PSD) of subthreshold oscillations in response to Poisson process driven conductance-based synaptic inputs by increasing the maximal synaptic conductance. For the numerical analysis, we used modified euler method with time step size Δt=0.05 *ms*. The PSD of membrane potential has been calculated using MATLAB. We used the fast Fourier transform (FFT) algorithm for the calculation of the PSD.

## 4. Results

### 4.1 Numerical results using 7D stellate cell model

The 7D full SC model shows subthreshold oscillation near 10 *Hz* and the amplitude of the PSD at the peak is 0.8871 when $I_{app} = -2.72 \mu A/cm^2$. For higher value of $I_{app}$, this model displays action potentials. Therefore, in our work, we set the control level (synaptic inputs are blocked) when $I_{app} = -2.72 \mu A/cm^2$ where the SC model displays only subthreshold oscillations. We calculated the PSD for voltage traces in response to synaptic inputs for different maximal synaptic conductance values. We compared the peak-amplitude in the PSD for voltage traces. To capture the experimental results [19], it is important to prevent the cell from firing since synaptic inputs induces the generation of action potentials. To achieve that, we adapt (decrease or increase) injected current.

#### 4.1.1 SC model captures the experimental results

Figure 2 shows that conductance-based synaptic inputs reduce the amplitude of subthreshold oscillations of membrane voltage while the current-based synaptic inputs do not change its peak at theta frequencies in the PSD. Figure 2A and B show synaptic conductance- and current-based inputs respectively. Figure 2C shows the PSD for membrane voltage under control, conductance- and current-based synaptic inputs. We used different maximal synaptic conductance for conductance- and current-based synaptic inputs since the amount of synaptic currents should be comparable (each synaptic input has the same mean value of synaptic input). For conductance-based synaptic input, we used $G_e = G_i = 0.05 \mu S/cm^2$ and $G_e = G_i = 0.5 \mu S/cm^2$ was used for current-based synaptic input. The difference between conductance- and current-based synaptic inputs is the amplitude of fluctuations in time (the mean values of each current are the same). The amplitude of conductance-based synaptic input is larger (almost 4 times) than the current-based synaptic input. But the effect of each currents is different. The conductance-based synaptic input significantly reduced the amplitude of subthreshold oscillations but current-based synaptic input does not alter the amplitude of subthreshold oscillations. These numerical results are in agreement with the experimental data [19]. Therefore, we hypothesize that the amplitude difference between conductance- and current-based synaptic current plays a key role in reducing or retaining the amplitude of subthreshold oscillations.

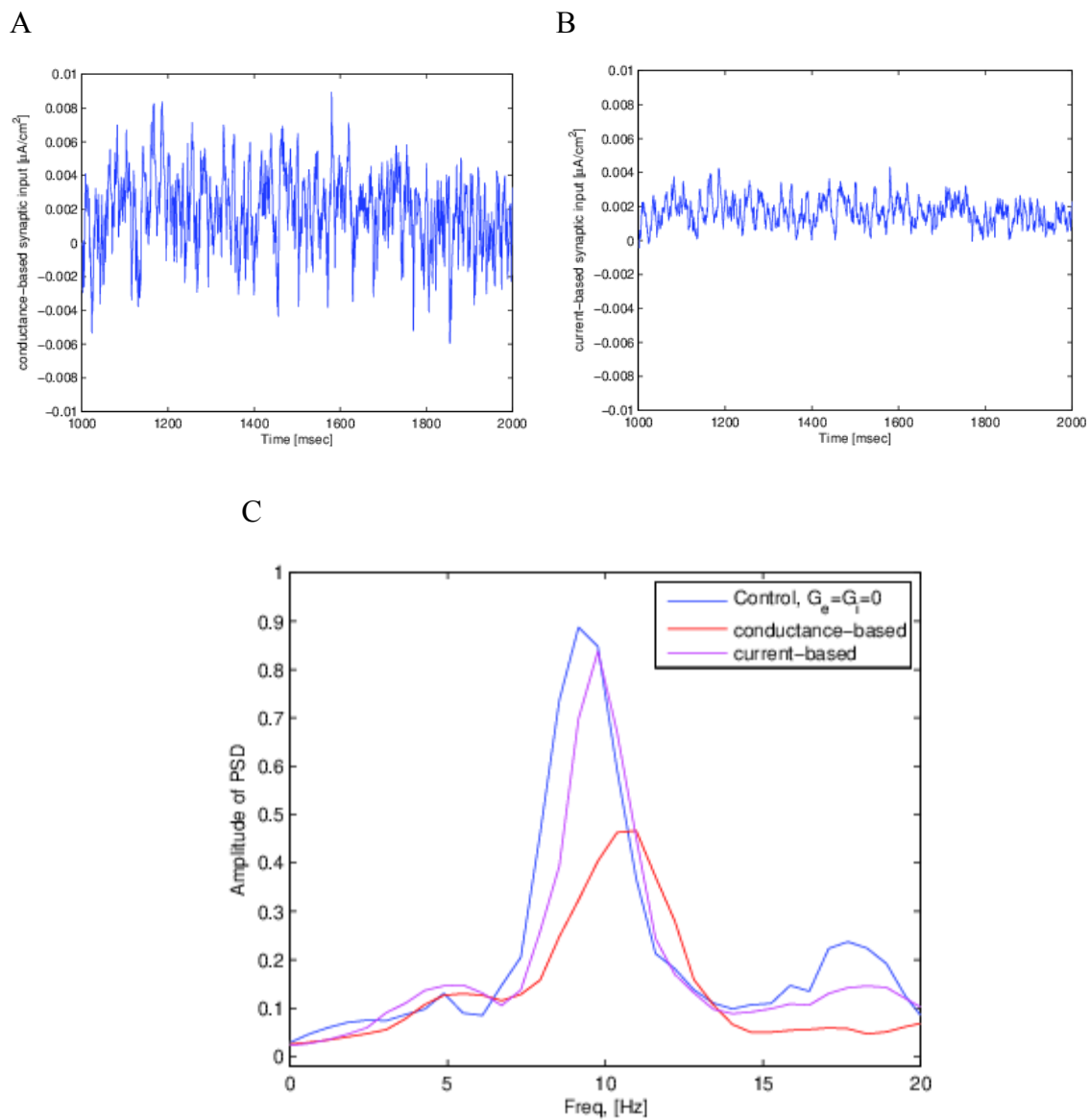

Figure 2: SC model captures the experimental result[19]. A and B show the conductance- and current-based synaptic inputs respectively. C shows the PSD for membrane voltage under control, conductance- and current-based synaptic inputs. We used the maximal synaptic conductance: $G_e = G_i = 0.05 \mu S/cm^2$

## 4.1.2 The effect of maximal synaptic conductance on the subthreshold activity under conductance-based synaptic inputs

Passive properties of cells act as low pass filter: i.e., they reduce the amplitude and frequency preference of subthreshold oscillations. Conductance-based synaptic inputs have been shown to produce a similar effect [19, 28]. An interesting question arises under what conditions these inputs behave like a leak current on the SCs. In this section, we investigated the effect of maximal synaptic conductance on the SC subthreshold activity under conductance-based synaptic inputs. Figure 3 shows the PSD for SC's subthreshold oscillations in respond to conductance-based synaptic inputs at various maximal synaptic conductance values. The peak-amplitude in the PSD for subthreshold oscillations for maximal synaptic conductance ranged from 0 to 0.15 $\mu S/cm^2$ is below the peak-amplitude at the control case but it starts increasing for higher values. This result is in contrast to the experimental results and suggests that the effect of conductance-based synaptic inputs does not behave like a leak current on SCs for higher maximal synaptic conductance values.

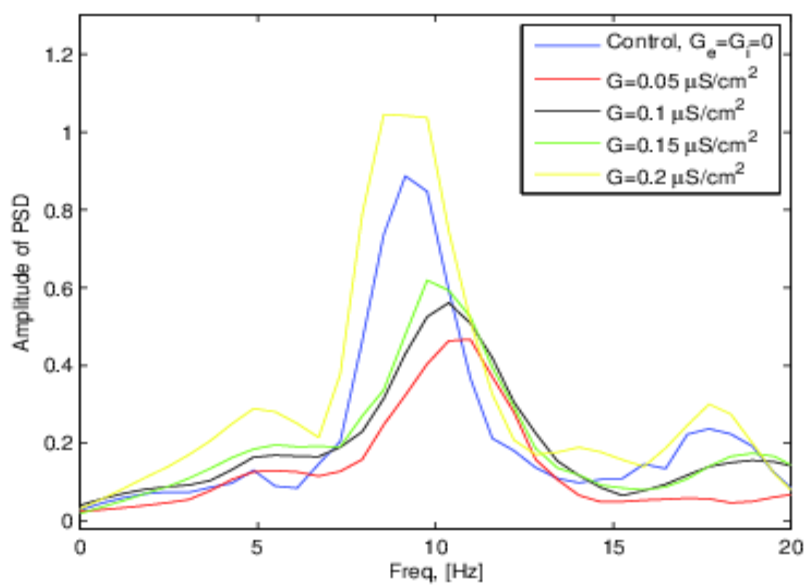

Figure 3: The PSD for voltage traces in response to conductance-based synaptic inputs at various maximal synaptic conductance values ($G_e = G_i = 0, 0.05, , 0.1, 0.15$ and $0.2$ $\mu S/cm^2$). The peak-amplitude in the PSD for subthreshold oscillations for maximal synaptic conductance ranged from 0 to 0.15 $\mu S/cm^2$ is below the peak-amplitude at the control case but it starts increasing for higher value.

Figure 4 shows conductance-based synaptic inputs with respect to time for various maximal synaptic conductance values ($G_{syn} = G_e = G_i = 0.05, 0.1, 0.15$ and $0.2$). Increasing $G_{syn}$ results in increasing the fluctuations of conductance-based synaptic inputs ($I_{syn}$). In order to understand why conductance-based synaptic inputs affect differently on the SC subthreshold activity with various $G_{syn}$, we will introduce $I_{S,L}$ which adapts leak current to include synaptic effects in next section.

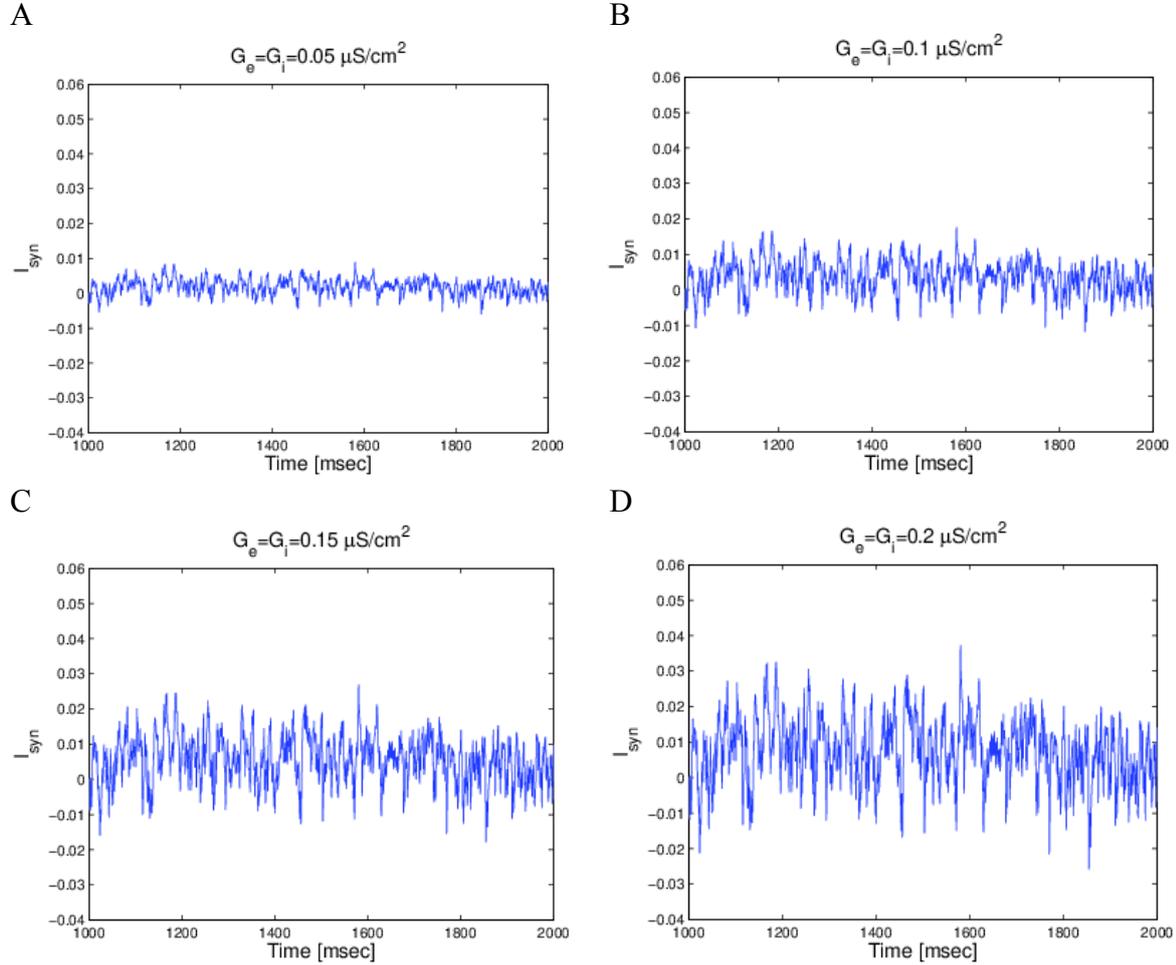

Figure 4: Conductance-based synaptic input ($I_{syn}$) with different values of maximal synaptic conductance ($G_{syn} = G_e = G_i$). Increasing $G_{syn}$ results in amplifying fluctuations of $I_{syn}$.

### 4.1.3 Modifying synaptic current with constant value of $S_e$ and $S_i$

The term $I_{syn}$ in equation (13) consists of $I_E$ and $I_I$, excitatory and inhibitory synaptic inputs respectively. Thus $I_{syn}$ in equation (14) and (15) can be rewritten in the following way

$$I_{S,L} = (G_e S_e + G_i S_i)(V - \frac{G_e S_e E_e + G_i S_i E_i}{G_e S_e + G_i S_i}) \qquad (16)$$

For constant values of $S_e$ and $S_i$ in Eqn. (16), $I_{S,L}$ is a passive current with a similar behavior as $I_L$ (leak current). The effective maximal conductance and reversal potential of $I_{S,L}$ are

$$G_e S_e + G_i S_i \quad \text{and} \quad \frac{G_e S_e E_e + G_i S_i E_i}{G_e S_e + G_i S_i}$$

The current-balance equation for $I_{S,L}$ is given by

$$C\frac{dV}{dt} = I_{app} - I_{Na} - I_K - I_L - I_h - I_p - I_{S,L} \qquad (17)$$

### 4.1.4 The effect of maximal synaptic conductance on the stellate cell's subthreshold activity under the presence of $I_{S,L}$

Figure 5 shows the PSD for subthreshold oscillations in respond to $I_{S,L}$ with $G_e$ and $G_i$ = 0, 0.05, 0.1, 0.15 and 0.2. In this simulation, we observed that $I_{S,L}$ reduced the peak-amplitude in the PSD by increasing $G_e$ and $G_i$. For $G_e = G_i = 0.05$, $I_{S,L}$ reduce the peak-amplitude without altering peak frequency but this current not only reduce peak-amplitude in the PSD for subthreshold oscillations but also alter peak frequency to higher frequency for higher value of maximal synaptic conductance. These results indicate that $I_{S,L}$ reduce subthreshold oscillations and that change the properties of subthreshold resonance. These are consistent to the previous results [19].

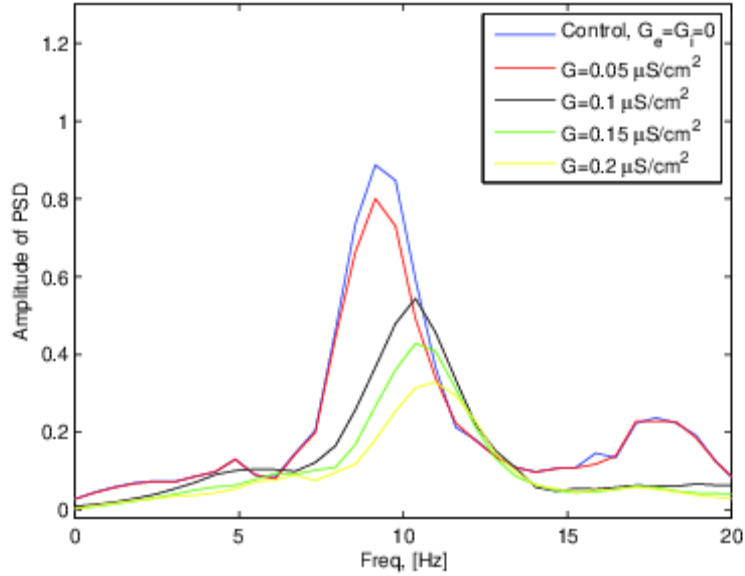

Figure 5: The PSD for subthreshold oscillations in respond to $I_{S,L}$ with $G_e = G_i$ = 0, 0.05, 0.1, 0.15 and 0.2. Adapting leak current, $I_{S,L}$, reduced the peak-amplitude in the PSD by increasing maximal synaptic conductance. For $G_e = G_i = 0.05$, $I_{S,L}$ reduce the peak-amplitude without altering peak frequency but this current not only reduce peak-amplitude in the PSD for subthreshold oscillations but also alter peak frequency to higher frequency for higher value of maximal synaptic conductance.

Figure 6 shows the activity of $I_{S,L}$ with respect to time for various maximal synaptic conductance values which were used in Poisson synaptic input case. In Poisson-driven conductance-based synaptic input, increasing $G_e$ and $G_i$ results in increasing the mean value of $I_{S,L}$. This result is different from Poisson case where increasing maximal synaptic conductance results in amplifying the fluctuations conductance-based synaptic inputs. This implies that SC receives inhibition from $I_{S,L}$, which cause the amplitude of subthreshold oscillations to reduce.

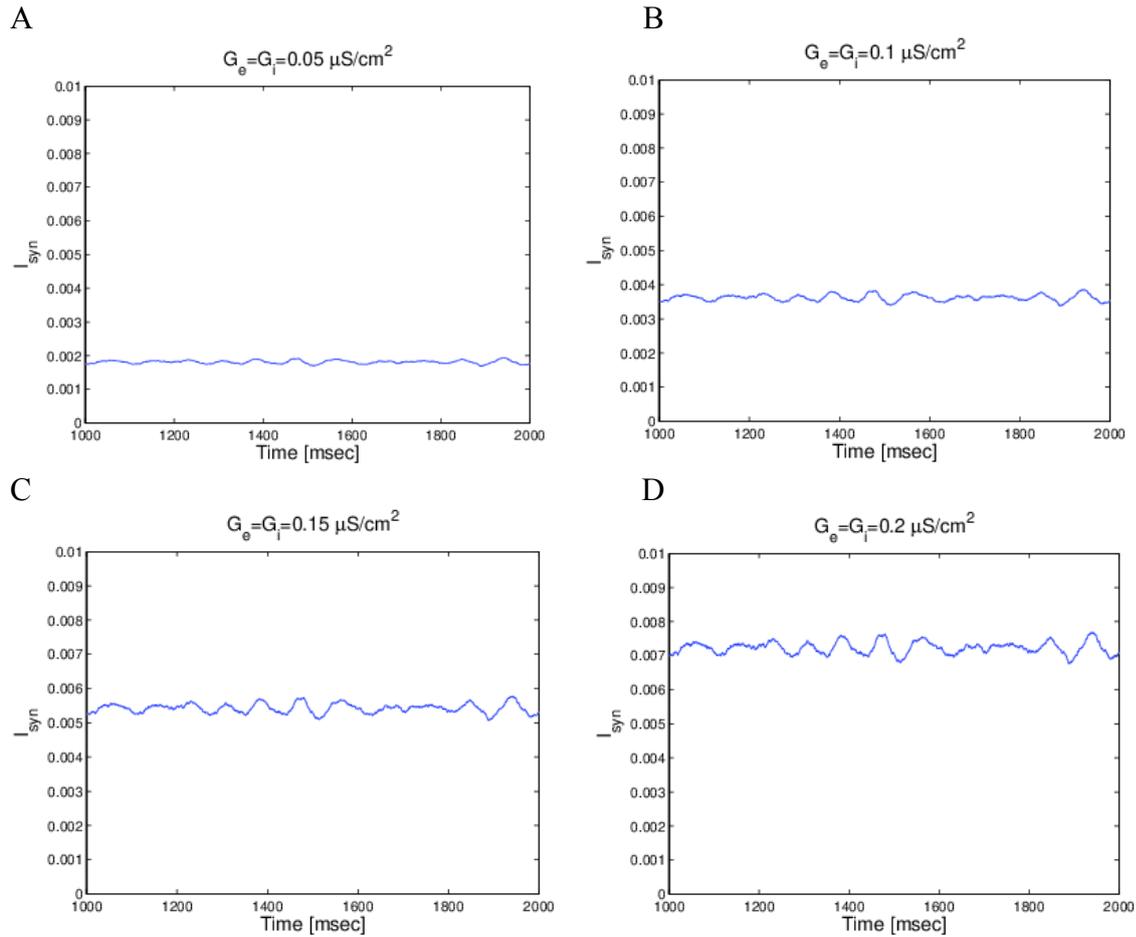

Figure 6: The activity of $I_{S,L}$ with respect to time for various maximal synaptic conductance values which were used in Poisson synaptic input case. These results display that increasing maximal synaptic conductance results in increasing the mean value of $I_{S,L}$

### 4.1.5 Different effects of $S_e$ and $S_i$ under Poisson driven conductance-based synaptic input and synaptic leak current

In previous section, we discussed the effect of pre-synaptic activity under both Poisson driven conductance-based synaptic inputs and $I_{S,L}$. Also, we studied changes of membrane potential responded to both synaptic inputs with various maximal synaptic conductance ($G_e = G_i =0$, 0.1, 0.15 and 0.2 $\mu S/cm^2$). Here, we investigate that the effect of $G_e$ and $G_i$ on the peak-amplitude in the PSD for subthreshold oscillations. We varied systemically $G_e$ and $G_i$ in both Poisson conductance-based synaptic inputs and $I_{S,L}$ (Fig. 7). In Poisson case, the peak-amplitude in the PSD first decreases but starts increasing (showing monotonic decreasing but changed the direction to monotonic increasing at some critical value ($\approx 0.5\ \mu S/cm^2$). Whereas, in the $I_{S,L}$ case, the numerical simulations exhibit standard behavior of the leak current ($I_L$). From these graphs, it can be thought that what biophysical parameter or mechanism produces the non-monotonic behavior in Poisson conductance-based inputs: large fluctuations of conductance-based synaptic inputs may produce non-monotonic behavior in peak-amplitude in the PSD.

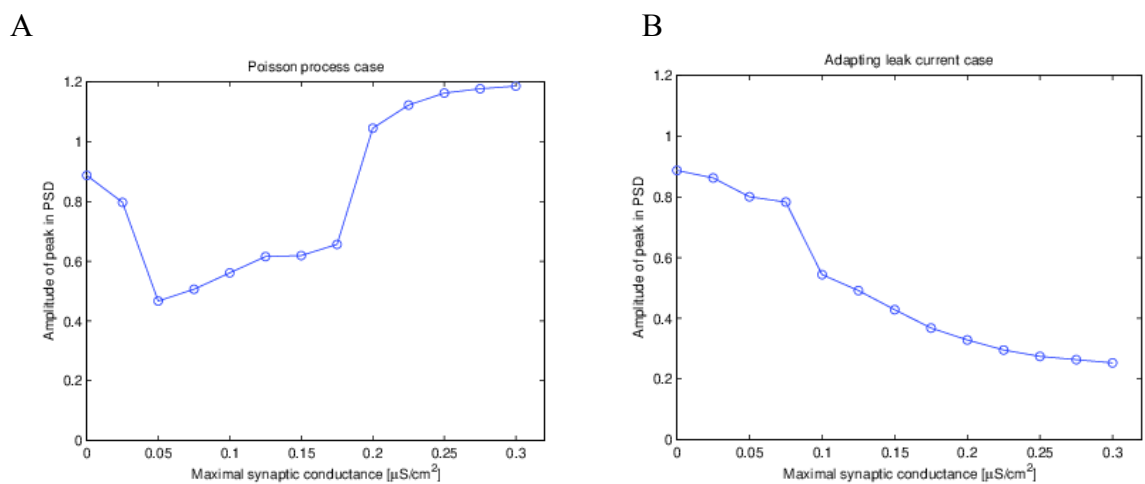

Figure 7: Systemically increasing maximal synaptic conductance shows different effects in both conductance-based synaptic input and $I_{S,L}$. A is the case of conductance-based synaptic input and B is the case of $I_{S,L}$ in terms of $G_{syn}$. For conductance-based synaptic input driven by Poisson process, the peak in the PSD first decreased and then started increasing at the critical value of maximal synaptic conductance while the peak in the PSD kept decreasing with increasing $G_e$ and $G_i$ in the $I_{S,L}$. case.

## 4.2 Numerical results using reduced 3D stellate cell model

In this section, we examined the effect of both conductance- and current-based synaptic currents (driven Poisson process) using the reduced 3D SC model [11]. We used the same parameters used in 7D SC model. The goal of this section is that the reduced 3D SC model is appropriate to reproduce the phenomena observed in 7D SC model and understand the evolution of dynamical system underlying reducing and amplifying the subthreshold oscillation under conductance-based synaptic inputs using phase-space analysis.

### 4.2.1 The reduced stellate cell model captures the experimental results

We set the control level ($I_{app} = -2.59 \mu A/cm^2$, no synaptic inputs) where 3D SC model exhibits only subthreshold oscillations. For the higher value of $I_{app}$, this model starts generating action potential. And we calculated the PSD of membrane potential under control, conductance-, and current-based synaptic current. We used the same excitatory and inhibitory synaptic protocols used in 7D SC model. We used $G_e = G_i = 0.05\ \mu S/cm^2$ for conductance-based synaptic inputs and $G_e = G_i = 0.5\ \mu S/cm^2$ for current-based synaptic inputs. In Figure 8, we show graphs of conductance- and current-based synaptic inputs

(Figure 8A and B, respectively) and the PSD for subthreshold oscillations in respond to no synaptic input (control), conductance-, and current- based synaptic inputs (Figure 8C).

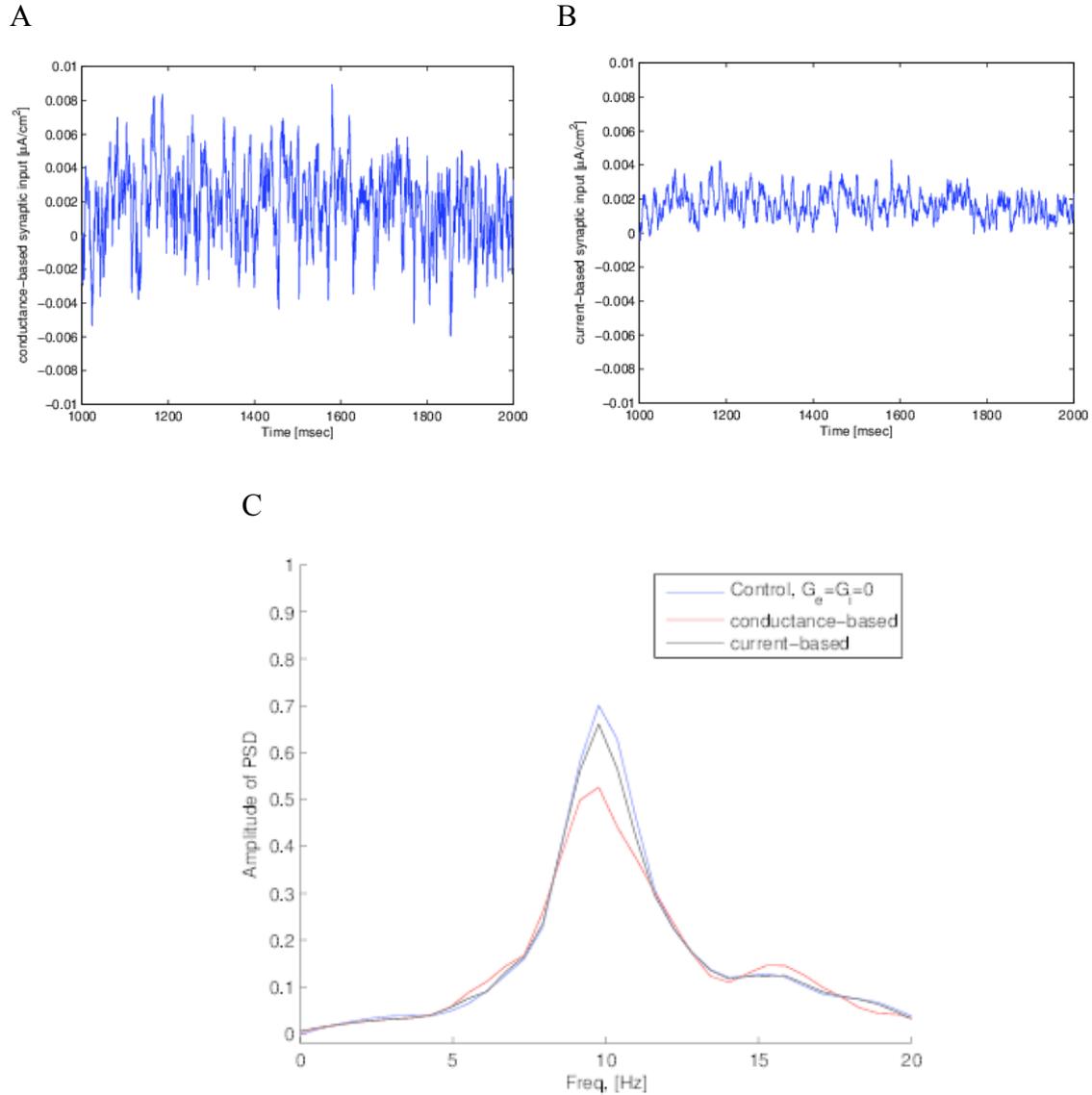

Figure 8: The reduced 3D SC model captures the experimental result[19]. A and B show the conductance- and current-based synaptic inputs respectively. C shows the PSD for membrane voltage under control, conductance- and current-based synaptic inputs. We used the maximal synaptic conductance: $G_e = G_i = 0.05\ \mu S/cm^2$ for conductance-based synaptic inputs and $G_e = G_i = 0.5\ \mu S/cm^2$ for current-based synaptic inputs

### 4.2.2 The effect of $G_e$ and $G_i$ on stellate cell under Poisson process and synaptic leak current

In this section, we studied the effect of $G_e$ and $G_i$ on SC under conductance-based synaptic input and adapting leak current ($I_{S,L}$). We measured the PSD for subthreshold oscillations in respond to both conductance-based synaptic current (Figure 9A) and adapting leak current

(Figure 9B) for $G_e = G_i$ = 0, 0.05, 0.15, and 0.2 $\mu S/cm^2$. Figure 10 shows the amplitude-peak of the PSD for subthreshold oscillations as a function of maximal synaptic conductance. When we systemically increased the maximal synaptic conductance from 0 to 0.3, the amplitude-peak in the PSD first decreased but started increasing at $G_e = G_i = 0.075$ $\mu S/cm^2$ in conductance-based synaptic input but the peak in the PSD kept decreasing in synaptic leak current. These numerical results are consistent to what we observed in 7D SC model.

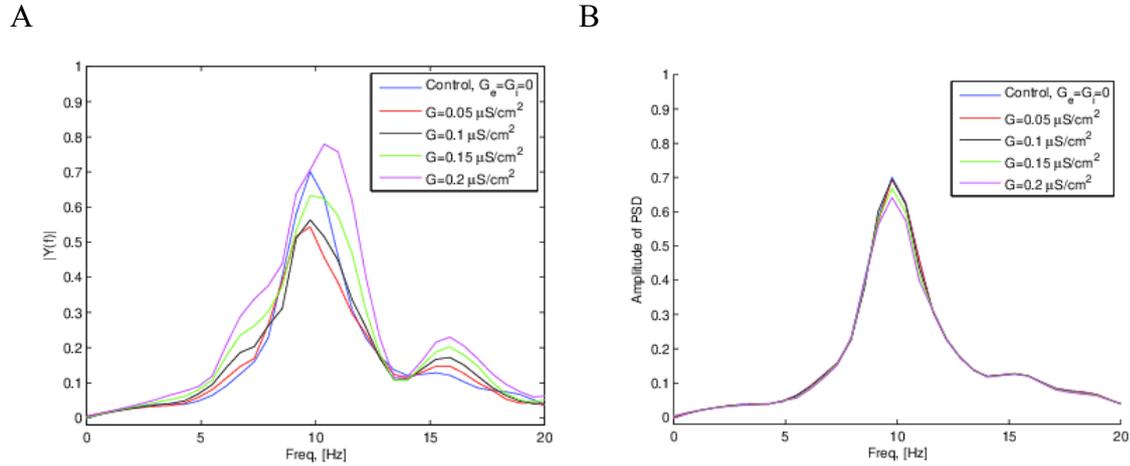

Figure 9: The PSD in 3D SC model changed by changing $G_e$ and $G_i$,. Here we used $G_e = G_i$ = 0, 0.05, 0.1, 0.15, and 0.2 μS/cm2 . The PSD under conductance-based synaptic input is shown in A and the PSD under synaptic leak current is shown in (B).

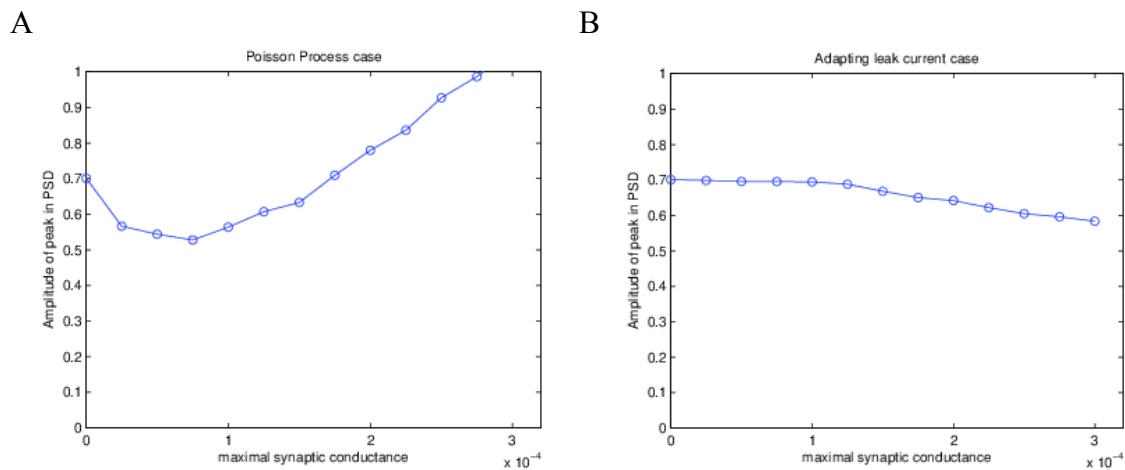

Figure 10: 3D SC model. Systemically increasing the maximal synaptic conductance shows different effects in both conductance-based synaptic input and synaptic leak current. The right panel (A) is the case of conductance-based synaptic input and the left panel (B) is the case of adapting leak current. For conductance-based synaptic input, the peak in the PSD first decreased and then started increasing at the critical value of maximal synaptic conductance while the peak in the PSD kept decreasing with increasing $G_e$ and $G_i$ for adapting leak current.

# 5. Conclusion and Disscusion

We studied the effects of conductance- and current-based synaptic inputs on a medial entorhinal cortex layer II stellate cell's subthreshold activity. Our numerical results are in agreement with the experimental results [19] : conductance-based synaptic inputs not only reduce the amplitude-peak in the PSD for corresponding subthreshold oscillations to these input but also alter subthreshold resonance properties on the SC while current-based synaptic input retains its peak at the theta frequencies. Previous experiment study [19] has suggested that these inputs behave like a leak current. However, we found that conductance-based synaptic inputs do not always act as a leak current. These results differ from the previous study. We found maximal synaptic conductance ranges where the behavior of conductance-based synaptic inputs changes at a critical values.

In this paper, we would argue that the effect of conductance-based synaptic inputs (driven by Poisson process) is shown leaky behavior for small enough maximal synaptic conductance value ($G_e$ and $G_i$) but for the higher value of $G_e$ and $G_i$, those currents do not behave like a leak current. Furthermore, we studied the effect of maximal synaptic conductance on both conductance-based synaptic and adapting leak current ($I_{S,L}$). From the numerical simulations, we suggest that the fluctuations of presynaptic activity may play an important role in reducing or amplifying the amplitude of subthreshold oscillations. This suggestion can be supported by the effect of current-based synaptic inputs for large enough value of maximal synaptic conductance. Figure 11 shows the amplitude-peak in the PSD for subthreshold oscillations as a function of maximal synaptic conductance under current-based synaptic inputs. In this simulation, we increased maximal synaptic conductance up to $0.005\ \mu S/cm^2$ . The amplitude-peak in the PSD as a function of maximal synaptic conductance behaves like a leak current in this regime.

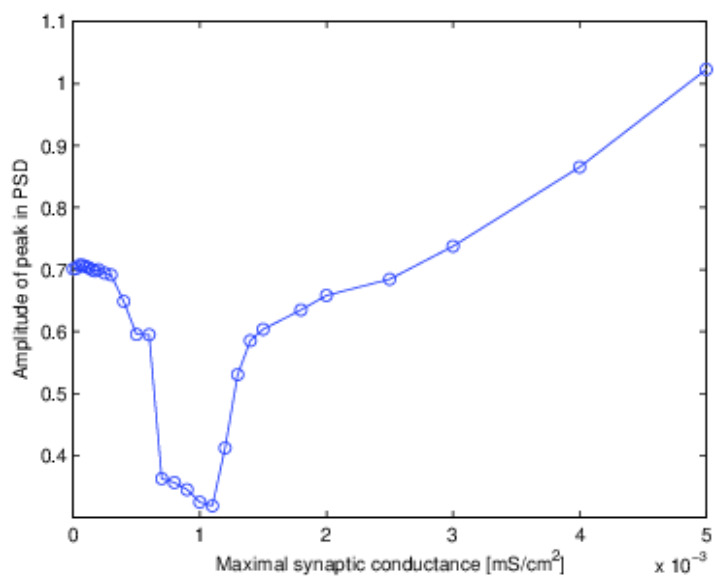

Figure 11: 3D SC model. Systemically increasing the maximal synaptic conductance under current-based synaptic inputs shows the same non-monotonic behavior in amplitude-peak : the amplitude-peak in PSD first decreases and then starts increasing at $G_e = G_i = 0.115\ \mu S/cm^2$.

# 6. Appendix

The definitions of the functions defining $x_\infty(V)$ and $\tau_x(V)$

$\alpha_m(V) = -0.01(V + 23)/(e^{-0.1(V+23)} - 1)$

$\beta_m(V) = 4e^{-(V+48)/18}$

$\alpha_h(V) = -0.07e^{-(V+37)/20}$

$\beta_h(V) = 1/(e^{-(0.1(V+7))} + 1)$

$\alpha_n(V) = -0.01(V + 27)/(e^{-0.1(V+27)} - 1)$

$\beta_n(V) = 0.125e^{-(V+37)/80}$

$\alpha_p(V) = 1/(9.15\left(1 + e^{-\frac{(V+38)}{6.5}}\right))$

$\beta_p(V) = e^{-\frac{(V+38)}{6.5}}/(0.15(1 + e^{-\frac{(V+38)}{6.5}}))$

$r_{f,\infty}(V) = 1/(1 + e^{-(V+79.2)/9.78})$

$\tau_{r_f}(V) = 0.51/(e^{\frac{(V-1.7)}{10}} + e^{\frac{-(V+340)}{52}} + 1)$

$r_{s,\infty}(V) = 1/(1 + e^{-(V+2.83)/15.9})^{58}$

$\tau_{r_s}(V) = 5.6/(e^{\frac{(V-1.7)}{14}} + e^{\frac{-(V+260)}{43}} + 1)$

$\tau_p(V) = 0.15$

$p_\infty(V) = 1/(1 + e^{-\frac{(V+38)}{6.5}})$